\documentclass[prb,floatfix,superscriptaddress,twocolumn,aps,showpacs,preprintnumbers,amsmath,amssymb]{revtex4}
\tighten
\renewcommand{\narrowtext} 
{\begin{multicols}{2}\global\columnwidth20.5pc} 
\newcommand{\be}{\begin{equation}}
\newcommand{\ee}{\end{equation}}
\newcommand{\bea}{\begin{eqnarray}}
\newcommand{\eea}{\end{eqnarray}}
\newcommand{\TK}{T_\text{K}}
\newcommand{\br}{{\bf r}}
\newcommand{\bk}{{\bf k}}
\newcommand{\ci}{\frak{i}}

\newcommand{\ed}{\epsilon_\text{d}}
\newcommand{\eds}{\epsilon_\text{d}^{*}}
\newcommand{\rqp}{\mathfrak{r}}

\newcommand{\fp}{\mathfrak{p}}

\usepackage{wasysym}
\usepackage{graphicx}
\usepackage{bm}

\newcommand{\ColorOnline}{(Color online) }
\usepackage{ulem}
\usepackage{color}
\definecolor{darkred}{rgb}{0.6,0,0}
\definecolor{red}{rgb}{1,0,0}

\begin{document} 
\preprint{INT/Evers-n}

\title{Broadening of the Derivative Discontinuity in Density Functional Theory}

\author{F. Evers}
\affiliation{
  Institute of Nanotechnology,
  Karlsruhe Institute of Technology, D-76021 Karlsruhe, Germany}
\affiliation{Institut f\"ur Theorie der Kondensierten Materie and Center of Functional Nanostructures, 
  Karlsruhe Institute of Technology, D-76131 Karlsruhe, Germany}
\email{ferdinand.evers@kit.edu}
\author{P. Schmitteckert}
\affiliation{
  Institute of Nanotechnology,
  Karlsruhe Institute of Technology, D-76021 Karlsruhe, Germany}
\email{peter.schmitteckert@kit.edu}

\date{\today}

\begin{abstract}
We clarify an important aspect of density functional theories,
the broadening of the derivative discontinuity (DD) in a quantum system,
with fluctuating particle number.
Our focus is on a correlated model system, the single level
quantum dot in the regime of the Coulomb blockade. We find that
the DD-broadening is controlled by the small parameter $\Gamma/U$, where
$\Gamma$ is the level broadening due to contacting and $U$
is a measure of the charging energy. Our analysis suggests,
that Kondoesque fluctuations have a tendency to increase the
DD-broadening, in our model by a factor of two. 
\end{abstract}
\maketitle

\paragraph*{Introduction.}
Over the years the density functional theory
(DFT)  developed into an
important tool to study transport properties of nano-systems and
single molecules. 
\cite{ke2008, gagliardi07,markussen10,dellAngela10,mishchenko10}
This development occured despite of the fact, 
that often the results are quantitatively 
sensitive to the approximations made for the 
{\it exchange correlation} (XC) functional, $V_\text{XC}[n]$,
underlying such calculations.  
\cite{ke2007,toher:146402,reimers03,evers03prb,sai:186810}
In principle, dc-transport calculations should combine either long-time
evolution of wavepackets or a KS-based quasistationary formalism, 
in both cases with special dynamical XC-functionals. 
\cite{stefanucci:195318, evers03prb, kurth:035308}
In practice, the available ground state functionals are being used. 

The neglect of dynamical correlations for simulation of dc-transport 
was justfied for spinless systems for which  a  Friedel-sum rule
holds. \cite{schmitteckert:086401}
In such systems, approximations to the XC-potential of the ground
state introduce the largest numerical error in the regime of 
{\it Coulomb blockade} (CB) where  the system
(``quantum dot'', QD)  is only weakly coupled to the electronic reservoirs
and the filling is close to an  integer.  Its signature is an addition 
energy, $U$, that largely exceeds the (single particle) level 
spacing of the QD.

In this Letter we exploit  the observation 
that  CB is intrinsically an {\it equilibrium} 
phenomenon even though it is mostly discussed in its effect on
transport measurements\cite{glazman05}; in a broader context it is a typical
manifestation of the reduced compressibility, $dn(\br)/d\mu$, of
repulsively interacting fermion gases. Therefore, it has a
reincarnation in XC-functionals of DFT where it appears as the 
{\it derivative discontinuity} (DD). 

Starting from the seminal work by Perdew et al. \cite{perdew82}, 
the DD was almost exclusively discussed in the limiting case of 
decoupled quantum dots, i.e. {\it closed systems}. 
There, the XC-functional jumps discontinuously when tuning the particle
number, $N$, of a closed system in its ground state through an integer value
\be
\Delta_\text{XC} = \lim_{{\delta N}\to 0}\left[ V_\text{XC}^{N+{\delta N}} -
  V_\text{XC}^{N-{\delta N}}\right].
\label{e1}
\ee
hence the name. In this context the DD often makes a quantitatively 
relevant contribution to the  band gap
\be
\Delta = \lim_{{\delta N}\to 0} \left[ \mu^{N+{\delta N}} - \mu^{N-{\delta N}} \right]. 
\label{e2}
\ee
where $\mu^{N}$ denotes the electrochemical potential of the $N$
particle system  which (up to a sign) equals 
the workfunction. \cite{rosselli06,sanchez08}
The relation between $\Delta$ and $\Delta_\text{XC}$ is easy to see. 
Due to Janak's theorem 
the energy of a KS-orbital, index $M$, 
in the  $N$-particle system, $\epsilon^{N}_{M}$
is related directly to the work functions: 
$\mu^{N-{\delta N}}{=}\epsilon^{N-{\delta N}}_{N}$, 
$\mu^{N+{\delta N}}{=}\epsilon^{N+{\delta N}}_{N+1}$. 
\cite{janak78,almbladth85} 
With Eq. (\ref{e2}) we conclude  
\be
\Delta =  \Delta_\text{KS} + \Delta_\text{XC},
\label{e3}
\ee
where $\Delta_\text{KS}$ is the energy spacing between the lowest
unoccupied ($M=N+1$, LUMO) and the highest
occupied ($M=N$, HOMO) KS states, 
$\Delta_\text{KS}{=} \epsilon^{N{-}{\delta N}}_{N{+}1}
{-} \epsilon^{N{-}{\delta N}}_{N}$,
and 
\be
\Delta_\text{XC} = \epsilon^{N{+}{\delta N}}_{N{+}1} {-}\epsilon^{N{-}{\delta N}}_{N{+}1}. 
\label{e3a}
\ee
It follows that the DD, $\Delta_\text{XC}$, accounts for the difference between the
bare, single particle gap, $\Delta_\text{KS}$, and the addition energy, 
$\Delta$, for supplying one more particle. 
This extra energy cost, $\Delta_\text{XC}$, related to the repulsive
interaction of fermions confined in a narrow region of space
is also the origin of the CB and incompressibility. 
The DD in closed $N$-particle systems and ways to include it into
approximate schemes have been a subject of
intense research, recently. \cite{kuemmel08}
In local or semi-local approximations of XC-functionals 
artifacts in the description of charge transfer and transport
processes arise, because the DD is not accounted for.  
\cite{kuemmel08,koentopp:121403,evers06}
\footnote{Most current density functionals rely upon local (LDA) and semi-local (GGA)
approximations\cite{fiolhaisBookPerdew}; they exhibit a continuous evolution with $N$
which implies that
the LUMO$^{N+{\delta N}}$ and the LUMO$^{N-{\delta N}}$ see essentially the same
  effective potential; therefore $\Delta^\text{LDA,GGA}_\text{XC}=0$.
  }

In {\it open systems} the understanding of the DD is still relatively
poorly developed. 
In particular, its fate in a situation with weakly coupled 
subsystems, e.g., QD and electronic/thermal
reservoirs,  has not yet been studied
systematically. 
This is our focus here. We 
investigate the equilibrium compressibility,
$dn(\br)/d\mu$,
in a generic model system, the {\it Anderson} 
(or single site Hubbard) model
with a repulsive on-site interaction $U$. 
The smearing of the discontinuity
(\ref{e1}), as a consequence of coupling to a reservoir, can 
be observed in simulations employing the density matrix
renormalization group (DMRG)\cite{White:PRL1992}.  Insight about parametrical 
dependency  is drawn from analytical results that also allow the 
construction of ``toy'' XC-functionals to study CB in DFT.

Specifically, we report the compressibility $dN/d\mu$  near integer
filling, $N=1$, in
three different temperature regimes. We summarize our findings. 
(i) For the isolated level
we have at nonzero temperatures ($\beta=1/T$) the exact result: 
$
\left.\frac{dv_\text{XC}}{dN}\right|_{N=1} {=} Te^{\beta U/2}{+}T{-}U/2.
$
The derivative is finite as $T>0$ and 
the discontinuity is broadened by temperature. A dimensionless measure
of this smearing is provided by the number of particles
$\delta N$ that need to flow into the level in order to fascilitate
the increase (``jump'')  of $\Delta(N)$ by $\Delta_\text{XC}\approx U$: 
\be
\delta N = U \left. dN/d\mu \right|_{N{=}1} = 
\beta U (e^{\beta U/2}+1)^{-1} \qquad T_\Gamma\apprle T
\label{e5}
\ee
In this perspective, the DD is the statement that 
in the zero temperature limit the amount of particles needed to drive
the jump becomes arbitrarily small. 
(ii) In the presence of a weak coupling to an electronic reservoir, the
single level acquires a width $\Gamma$. Below a certain cross-over
temperature, $T_\Gamma ~ \Gamma$, we witness that  $\delta N$ stops to decrease and 
the lifetime broadening leads to an intermediate saturation; we obtain
\be
\label{e6}
\delta N \approx 4\Gamma/\pi U(1+{\cal O}(\Gamma/U))  \qquad
\TK\ll T \apprle T_\Gamma 
\ee
(iii) At even lower temperatures, below the Kondo scale, $\TK$, 
where the Abrikosov-Suhl resonance if fully developed, 
we cite an exact asymptotic result: 
\be
\label{e7}
\delta N = 8\Gamma/\pi U ( 1+{\cal O}(\Gamma/U)) \qquad T\ll \TK
\ee
In cases (ii) and (iii) model XC-functionals are given, that
reproduce the CB-features on a qualitative level. 

\paragraph*{Anderson model.}
The Anderson model\cite{hewsonBook} 
 describes a single level QD
coupled to a reservoir (${\mathcal R}$): 
\be
\hat H = \hat H_\text{QD} + \hat H_{\cal R} + V\sum_{\sigma=\uparrow,\downarrow} 
\sum_{\bk} \left( 
c^\dagger_{\sigma\bk} d_\sigma + d^\dagger_\sigma c_{\sigma\bk} \right), 
\ee
where 
$\hat H_{\cal
  R}{=}\sum_{\sigma\bk}\epsilon_{\bk}c^\dagger_{\sigma\bk}c_{\sigma\bk}$,
$\epsilon_{\bk}=-2t\cos(ka)$ ($a$: lattice spacing) 
and  in the presence of spin-rotational invariance 
($\hat n_\sigma{=}d^\dagger_\sigma d_\sigma, 
\hat N{=}\hat n_\uparrow + \hat n_\downarrow$,
$\langle \hat n_\sigma\rangle{=}N/2$):
$
\hat H_\text{QD} {=}\ed \hat N+ U \hat n_\uparrow \hat
n_\downarrow 
$
The observable of interest in DFT is the local density and its
variation with exernal parameters, e.g. $\ed$ and $\mu$: 
$N(\mu)$ and $dN/d\mu$ especially near integer
fillings, $N=1$. The inverse,  $\mu(N)$, will then be 
related to the exchange-correlation potential on the QD via
\be
v_\text{XC}(N) = \mu(N) - UN/2 - \ed; 
\label{e10}
\ee
Hartree and on site potential here must be subtracted.  
To calculate physical observables in the presence of reservoirs, 
thermal Green's functions provide a convenient formalism. 
For the thermal occupation numbers we
have quite generally
\be
N/2=\langle \hat n_\sigma\rangle = 
T \sum_{m}{\mathcal
  G}_\sigma(\ci \omega_m)
 = \int \! dE f_E {\cal A}_\sigma(E)
\label{e11}
\ee
where we have introduced the spectral function 
${\cal A}_\sigma(E) = (-1/\pi) \sum_\sigma \Im {\mathcal G}_\sigma(E)$
and $f_\epsilon
=(e^{\beta(\epsilon-\mu)}+1)^{-1}$.

\paragraph{Thermal coupling.} 
In the absense of particle fluctuations, $V=0$, the spectral function
can be calculated exactly. \cite{hewsonBook}
It is given by  ($\bar\sigma=-\sigma$)
\be
\label{e12}
{\cal A}_\sigma(E) = (1-\langle n_{\bar\sigma}\rangle)\delta(E-\ed) +
\langle n_{\bar\sigma}\rangle \delta(E-(\ed+U))
\ee
The two ``Hubbard bands'' are 
reflecting the energy cost, $U$, for adding the 
second particle to the QD. 
Recalling (\ref{e12}) we obtain
\be
\label{e13}
N/2 = f_{\ed} (f_{\ed}+f_{2\mu-\ed-U})^{-1},
\ee
which implies that at integer filling, $N{=}1$, we have 
$\mu_1{=}\ed{+}U/2$. 
Recalling 
equation (\ref{e10}) we conclude $v_\text{XC}|_{N=1}=0$. 
By inverting (\ref{e13}) we obtain the general answer 
\be
\mu(N) = T\ln \left[ 
\frac{N{-}1}{2{-}N}e^{\beta U} + 
\frac{\mathfrak{a} e^{\beta U/2}}{2-N} 
\right] +\ed
\ee
and
\be
d\mu/dN = T(e^{\beta U}-1)\ (\mathfrak{a}e^{\beta U/2} -\mathfrak{a}^2)^{-1}. 
\ee
 with $\mathfrak{a}(\beta U,N)=\sqrt{1+(e^{\beta U}{-}1)(N{-}1)^2}$. 
It is implied that near integer filling
$d\mu/dN|_{N=1} = T+Te^{\beta U/2}$
and 
\be
\left.\frac{dv_\text{XC}}{dN}\right|_{N=1} = Te^{\beta U/2}+T-U/2
\ee
From this expression it is obvious how the DD
emerges: at any nonzero value of the interaction parameter $U$, 
the slope near $N=1$ diverges in the zero temperature limit. 
The divergency occurs in an exponential way because those fluctuations
in particle numbers that give $\mu$ a nonvanishing slope are
suppressed by a factor of $\exp{\beta U/2}$. 

The diverging slope can also be interpreted in the following way. At
low temperature and near integer filling a very small change in the local
particle number, $\delta N$, can increase the effective on-site potential by the
finite amount $U$: 
$
\delta N \equiv U dN/d\mu = \beta U (e^{\beta U/2}+1)^{-1} 
$
We have arrived at Eq. (\ref{e6}).

\paragraph{The quantum  limit: $T\to 0$.}
 In the presence of the nonvanishing coupling, $V>0$, the occupation
numbers  $\hat n_\sigma$ no longer commute
with the Hamiltonian, $\hat H$, and the Anderson model 
becomes nontrivial. Two essential changes occur. 
First, the Hubbard bands acquire a finite width, $\Gamma$. 
As a consequence, the decrease of $\delta N(\beta U)$ stops when 
$T$ falls below $\Gamma$. The residual density fluctuations 
near integer filling then are no longer controlled by thermal but 
by quantum fluctuations. The control parameter for the latter  is 
$\Gamma/U$; it measures the overlap of the Hubbard bands with the
Fermi-energy. 
Second, at lowest temperatures, $T\ll \TK$,  the spectral function acquires a 
third peak, the {\it Abrikosov-Suhl resonance}. 

\begin{figure}[ht]
\begin{center}
    \includegraphics[width=0.48\textwidth]{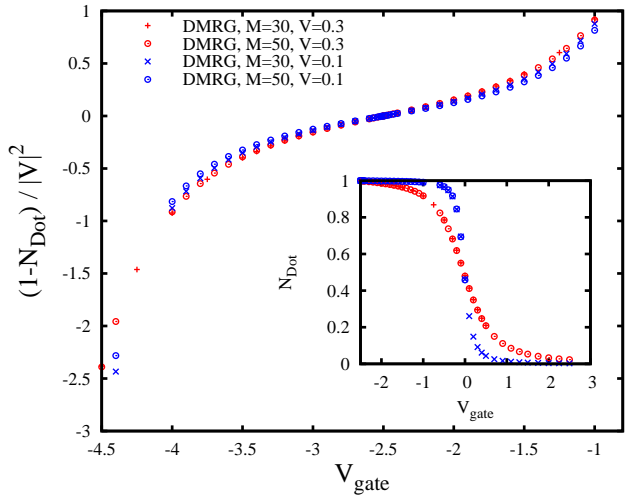}
\end{center}
     \caption{\ColorOnline The change of the dot filling rescaled by $|V|^{-2}$
          in the Coulomb blockade regime for
          an on-site repulsion of $U=5t$, a hybridization
          of $V=0.3t$ ($+$) and $V=0.1t$ ($\times$)
          for a $M=30$ site system obtained from a ground state DMRG. For comparison results
          for a 50 site calculation are also shown (red,blue $\circ$)
          emphasizing convergence with the system size.
          Data illustrates the scaling of $dN/d\mu$ with
          the hybridization $\Gamma\sim |V|^2$.
          Raw data is shown in the inset. 
	  Inset: dot occupation. }
   \label{fig:LDAvsDMRG}
\end{figure}

\paragraph{Intermediate temperatures: $\TK\ll T \apprle\Gamma,U$.}
We first imagine that the Kondo temperature is by far the
smallest energy scale, in particular $\TK\ll T$, so that the
Kondo effect can be ignored. This is justified in effectively finite 
systems, like large molecules, where the
density of states (DoS) near the highest occupied molecular energy level is
not truely continuous. In this situation the presence of the
reservoirs is easily dealt with on a qualitative level by equipping 
the thermal Green's function 
Eq. (\ref{e11}) with a self energy, $\Sigma(\ci \omega_m)$. 
It determines the inverse lifetime
$\Gamma(E) = -\Im \Sigma(E)$ 
which evaluates to $\Gamma(E) = \pi |V|^2\rho_{\mathcal R}(E)$
for noninteracting reservoirs 
($\rho_{\mathcal R}(E)$:  reservoir DoS).
The qualitative effect can be studied in the simplest approximation
where we ignore the energy dependency of the self energy
(``wide band limit'').  
Then, the spectral function for a given Hubbard subband 
takes the Lorentzian shape 
\be
{\cal L}(E) = \frac{1}{\pi}\frac{\Gamma}{E^2 + \Gamma^2}.
\label{e17}
\ee

As a consequence of the peak broadening, Eq. (\ref{e13}) generalizes, 
\be
\frac{N}{2} = \frac{F_{\eds}}{F_{\eds}+F_{2\mu-\eds-U}}, \quad 
F_u = \int_{-\infty}^{\infty} \! dE\ f_{E+u}  \ {\cal L}(E), 
\label{e18}
\ee
with the zero temperature limit
\be
F_u \mathop{=}_{T\to 0} 1/2+ \arctan{[(\mu-u)/\Gamma]}/\pi 
\ee
where $\eds=\ed+\Re\Sigma$.
Eq. (\ref{e18}) reveals that at $N=1$ we still have $\mu_1=\eds+U/2$ at
arbitrary $T,\Gamma$ values. 
The formulae (\ref{e18},\ref{e19}) combine into a transcendental equation for $\mu(N)$
 and by virtue of Eq. (\ref{e10}) into a functional $v_\text{XC}$. 

For calculating the $dN/d\mu$ we notice,
that $\mu$ enters $F_{\eds}$ and $F_{2\mu-\eds-U}$ with the opposite sign,
  implying  that the $\mu$ derivative of the denominator of
  (\ref{e18}) vanishes at $N=1$.  Hence we derive 
\be
\frac{dN}{d\mu} =F_{\eds}^{-1}
\left.\frac{dF_{\eds}}{d\mu}\right|_{N=1}
\mathop{=}_{T\to 0}
\frac{{\cal L}(U/2)}{1/2
  + \arctan{(U/2\Gamma)}/\pi}
\label{e19}
\ee
which implies 
\be
\label{e21}
\delta N = \frac{1}{\pi}\frac{2\Gamma/U}{1+(2\Gamma/U)^2} \frac{2}{1/2
  + \arctan{(U/2\Gamma)/\pi}}.
\ee
Eq. (\ref{e6}) follows via expansion in $\Gamma/U$. 

\paragraph{The Kondo limit: $T\ll \TK$.}
When the temperature decreases down to 
the Kondo scale, $T \sim \TK$,
\be
\TK = \mathfrak{c}
\sqrt{U\Gamma}e^{-\pi|\mu-\eds||\mu-\eds-U|/2U\Gamma}, 
\quad \eds \apprle \mu  \apprle \eds + U 
\ee
the Abrikosov-Suhl (AS) resonance starts to build up. \cite{hewsonBook} 
($D$: conduction electron bandwidth, $\mathfrak{c}\approx
(D^2/|\mu-\eds||\mu-\eds-U|)^{1/2}$; 
$\mathfrak{c}\approx 0.29$ in the wide band limit, $D\to\infty$.)
When it is fully developed, 
$T\ll \TK$, its shape is roughly Lorentzian, 
$
{\cal A}_\text{AS}(E) {\approx} (1/\pi\Gamma) 
T^2_\text{K}/((E{-}\mu)^2 {+} \TK^2)^{-1}, 
$
and it adds a third resonance to the spectral function 
\be
\label{e22}
{\cal A}(E) \approx \fp \sum_{\sigma}{\cal A}_\sigma(E) +\fp {\cal A}_\text{AS}(E)
\ee
with a normalizing coefficient, $\mathfrak{p}{=}1/(1{+}\TK/2\Gamma)$. 
As written, Eq. (\ref{e22}) has an artificial feature in the sense
that the peak values at all resonances coincide: $1/\pi\Gamma$.  
A slightly more accurate representation incorporates a change in the
shape of the Hubbard bands in the Kondo regime which we here 
account for by replacing the original width of ${\cal A}_\sigma$ with 
another one, $\Gamma \to \Gamma^\text{K}$, specified below. 
 With this caveat we have
($
\rqp= \int dE f_E {\cal A}_\text{AS}(E)=\TK/2\Gamma
$): 
\be
N  {=}\frac{2\ \fp F_{\eds} + \fp \rqp}{1{-}\fp {+}
  \fp(F_{\eds}{+}F_{2\mu-\eds-U})} 
  {=}
\frac{2 F_{\eds} +
\rqp}{F_{\eds}{+}F_{2\mu-\eds-U}{+}\rqp}. 
\label{e26}
\ee

Eq. (\ref{e26}) captures qualitative features of $N(\mu)$
and it can be used to construct an LDA
for a Kondo-system. Eq. (\ref{e26}) 
suggests that the impact of the AS-resonance on 
the compressibility is small as $\TK/\Gamma$. 
The main impact of Kondoesque fluctuations 
comes here from the renormalization of the shape of the Hubbard 
peaks. Indeed, the exact compressibility known from 
Bethe-Ansatz calculations\cite{hewsonBook}
reads in the limit of large $U$: 
\be
\left.\frac{dN}{d\mu}\right|_{N=1} = \frac{8\Gamma}{\pi U^2}
\left( 1 - \frac{6}{\pi} \frac{2\Gamma}{U}+
\ldots
\right).  
\label{e27}
\ee 
implying Eq. (\ref{e7}). Analogous to Eq. (\ref{e6}), an estimate based on (\ref{e26}) would give 
$4\Gamma^\text{K}/\pi U^2$  which suggests $\Gamma^\text{K}\approx
2\Gamma$ when comparing with (\ref{e27}). 
We conclude that in the Kondo-regime $dN/d\mu$ 
should be enhanced by a factor $\sim 2$
reflecting a stronger tendency for
charge-fluctuations. 

{\paragraph{DMRG-calculation.}
  To illustrate our analytical arguments 
DMRG-calculations have been performed on systems including
$M=30,50$ sites.
The available system sizes
do not allow us to resolve the Kondo-scale, yet,
but the expected finite slope of $dN/d\mu$
at integer filling is clearly visible. 
The data also shows the collaps
on a single curve when rescaled by  $|V|^{-2}$.
}

\paragraph*{Discussion.}
Our survey of analytical results obtained in the symmetric
Anderson model suggests that the DD is
broadened in systems coupled to a reservoir:  the particle transfer, $\delta N$,
needed to shift the local XC-potential by the on-site interaction 
energy $\sim U$ is not infinitesimally small. 
Within the model considered, the particle transfer is a two-parameter
function, $\delta N(\beta U, \Gamma/U)$, that smoothly interpolates
between a high temperature and a low-temperature (``quantum'')
regime. 

This result has implications for model studies of quantum transport
within the framework of time dependent DFT. The importance of the DD
for such transport simulations has been emphasized in several recent 
works. \cite{toher:146402, koentopp:121403, kurth10} 
Our work implies, that effective functionals used in such simulations
should exhibit a parametric dependency on $\Gamma/U$. In particular,
only terms quadratic in the coupling, $V^2$, appear in single channel 
quantum transport. The ABALDA-functional proposed in Ref. [\onlinecite{kurth10}] 
does not adhere to this principle since it depends explicitly on the
parameter $U/V_\text{link}$.


Our  results also have implications for molecules, i.e. systems where
one subsystem couples weakly to a small number of other subsystems,
but not to a (macroscopic) reservoir, proper. 
In this situation, the spectral function ${\cal A}_\sigma$, Eq. (\ref{e12}),  will
translate into the local spectral function of the given subsystem: 
each Hubbard peak acquires a splitting indicative of the hybridization of
states with the environment.  Again, the amount of charge needed to
fill into the subsystem to drive $v_\text{XC}$ up by $U$ will not be
zero but rather reflect this hybridization induced substructure. 

Discussions with P. W\"olfle  and support by the CFN and SPP 1243 
are gratefully acknowledged. After completing and submitting
our work we became aware of independent research
by Bergfield, Liu, Burke
and Stafford.
\cite{bergfield11}.
Where they overlap,
their conclusions coincide with the ones presented
in this paper.


\bibliographystyle{apsrev}
\bibliography{./BibME/bibDft,./BibME/bibDftApplTransport,./BibME/bibInelastic,./BibME/bibOwnMolEl,./BibME/bibNegf,./BibME/bibTDFT,./BibME/books,./BibME/bibGeneral,./BibME/bibMethods}

\end{document}